\documentclass[twocolumn,prd,nofootinbib]{revtex4}
\usepackage{hyperref}
\usepackage{amsmath}
\usepackage{graphicx}
\usepackage{color}

\newcommand{\hidetosubmit}[1]{}
\newcommand\ForInternalReference[1]{}

\newcommand\qmstate[1]{\left|#1\right>}
\newcommand\qmstateKet[1]{\left<#1\right|}
\newcommand\qmstateproduct[2]{\left<#1|#2\right>}

\newcommand\localfile[0]{/Users/oshaughn/unixhome/}
\newcommand\localhost{localhost}
\newcommand\refsheetWebLink[1]{\href{http://\localhost/ReferenceSheets/#1.pdf}{pdf}}

\newcommand\refsheetDVILinkWithLabel[2]{\href{\localfile/NotesAndReferences/ReferenceSheets/#1.dvi}{#2}}

\newcommand\refsheetLinkWithLabelInline[2]{#2(\refsheetDVILinkWithLabel{#1}{local dvi}; \refsheetWebLink{#1})}

\newcommand\sampleDVILinkWithLabel[2]{\href{\localfile/NotesAndReferences/SampleProblems/#1.dvi}{#2}}

\newcommand\sampleLinkWithLabelInline[2]{#2(\sampleDVILinkWithLabel{#1}{local dvi})}

\newcommand\editremark[1]{ {\color{red} #1}}

\newcommand\nObs{{n}}

\newcommand\dparam{{d_{param}}}

\begin{document}
\title{Comparing compact binary parameter distributions  I: Methods} 
\author{R.\ O'Shaughnessy}
\affiliation{Center for Gravitation and Cosmology, University of Wisconsin-Milwaukee,
Milwaukee, WI 53211, USA}
\email{oshaughn@gravity.phys.uwm.edu}
\begin{abstract}
Being able to measure each merger's sky location, distance,  component masses, and conceivably spins,  ground-based
gravitational-wave detectors  will provide a extensive and detailed sample of coalescing compact binaries (CCBs) in the local and, with third-generation
detectors, distant universe.  
These measurements will distinguish between competing progenitor formation models. 
In this paper we develop practical tools to characterize the amount of experimentally accessible information available,
to distinguish  between two a priori progenitor models.
Using a simple time-independent model, we demonstrate the information content scales strongly with the number of
observations.   The exact scaling depends on how significantly mass distributions change between similar models.
We develop phenomenological diagnostics to estimate how many models can be distinguished, using first-generation and
future instruments.  
Finally, we emphasize that multi-observable distributions  can be fully exploited only with very precisely calibrated
detectors, search pipelines, parameter estimation, and Bayesian model inference.
\end{abstract}
\maketitle

\hidetosubmit{
\subsection{Theory: internal ref!}

\refsheetLinkWithLabelInline{ProbabilityAndMeasure/Probability/Continuous/cont_probability}{continuous probability}

\refinclude{math/ProbabilityAndMeasure/Probability/Continuous/review.GaussianInformation.tex}
}

\section{Introduction}

Ground-based gravitational-wave interferometric detectors like advanced LIGO and Virgo are expected to detect at least a
few coalescing compact binaries (CCBs) per year, based both on semi-empirical extrapolations of Milky Way binary pulsar
statistics  \cite{Chunglee-nsns-1,Chunglee-nsns-proceedings,psr-spinFb}
and on theoretical predictions both of isolated binary  \cite{2007PhR...442...75K,PSgrbs-popsyn,PSellipticals,popsyn-LowMetallicityImpact-Chris2008,popsyn-LowMetallicityImpact-Chris2010,popsyn-LowMetallicityImpact2-StarTrackRevised-2012}
and clustered evolution  \cite{PZMcM,clusters-2005,2008ApJ...676.1162S,2011MNRAS.416..133D}.  
Each detected waveform should reveal the sky location, distance, component masses, and conceivably even component
spins.  
Corrected for the known biases of gravitational-wave searches, %
the observed multi-property distribution can be compared statistically against any proposed model for CCBs.  In other words,
the set of binary mergers will help us choose between competing models for binary compact formation.

Comparison of models against data is well-explored in general \cite{book-BurnhamAnderson-ModelSelection,2011RvMP...83..943V}, in astronomy
\cite{2009MNRAS.398.1601F}, in stellar and binary evolution \cite{Chunglee-nsns-1,psr-spinFb,PSmoreconstraints,popsyn-constraint-StellarMassBHMassDistribution-Empirically-Farr2010,2004MNRAS.352.1372B,2003ApJ...589L..37B}, and  in
gravitational wave astrophysics \cite{2011PhRvD..83h2002D,2011PhRvD..84f2003C,gr-extensions-tests-Europeans2011}.   In brief, parameter and estimation and hypothesis testing relies on Bayes' theorem,
where a particular ``data'' (here, the set of all signals) is compared with a (possibly continuous) space of models:
\begin{eqnarray}
\label{eq:Bayes:NotationIntro}
p({\cal H}|d) = \frac{p(d|{\cal H}) p({\cal H})}{p(d)}
\end{eqnarray}
where $p(d|{\cal H})$ is the marginal probability for the data given ${\cal H}$; $p({\cal H}|d)$ is the posterior
probability given the data is observed; $p({\cal H})$ are our prior probabilities for the data; and  $p(d)$  is
self-consistently set to account for all models under consideration:
\begin{eqnarray} 
 p(d) &=& \int d{\cal H} p(d|{\cal H}) p({\cal H})
\end{eqnarray}
Models with  the highest posterior probability are favored; relative probabilities give ``odds ratios.''  
The above expressions are often expressed with coordinates for the model space, denoted henceforth by $\lambda$.
While generic and effective, Bayes' theorem's utility relies on our ability to efficiently examine all reasonable
scenarios and, for each, to accurately estimate posterior probabilities $p(d|{\cal H})$.  
All state-of-the-art  detailed source models for binary compact object formation are  computationally
expensive \cite{StarTrack2,1996AA...309..179P}.    %
Except for a handful of efforts
\cite{PSmoreconstraints,PSgrbs-popsyn,PSellipticals}, few large-scale parameter surveys have been attempted.  
Moreover, these Monte Carlo studies estimate posterior probabilities $p(d|{\cal H})$ with limited accuracy.   
Therefore, though computation-intensive (Markov-chain Monte Carlo) methods have been usefully applied to estimate
$p({\cal H}|d)$ for similarly complicated distributions, such as for the parameters of a \emph{single} binary merger \cite{2008ApJ...688L..61V,2009CQGra..26k4007R,gwastro-mergers-PE-Aylott-LIGOATest},  a complete blind survey of the binary evolution model space
remains computationally unfeasable.

Bayes theorem also does not interpret the information it provides: it only provides posteriors.   One way of
interpreting that information arises naturally, if the problem is nearly solved.    Eventually, many observations may tightly constrain
the parameters $\lambda$ of some hypothetical formation scenario to a small coordinate region surrounding an optimal
model $\lambda^*$.   In this limit, the posterior from a set of observations can be well-approximated by a narrowly peaked gaussian:
\begin{eqnarray}
p(\lambda|d) \propto \exp - \frac{\Gamma_{ab} (\lambda-\lambda^*)_a(\lambda-\lambda^*)_b}{2}
\end{eqnarray}
for some matrix $\Gamma_{ab}$.     In this well-understood limit, each subsequent further observation
both refines the optimal parameters and narrows the likelihood (i.e., increases $\Gamma$).
The impact of any one proposed measurement (or model variation) can be computed perturbatively.
While asymptotically useful, at this stage we have neither an adequately accurate model for binary evolution nor a good
handle on what model parameters nature prefers.
Clearly more robust diagnostics are required.

Realistically,  many  models predict a similar rate and mass distribution as any we could conceivably reconstruct from
the first gravitational wave detections.   To best exploit the available information gravitational wave signals provide,
we will use both the number and properties of each gravitational wave signal.  But how?
Equally realistically,  the number and properties predicted by neighboring distributions depends on the reference point
where we identify neighbors: the answer we seek depends on what nature provides.
How can we  assess  how much and what information each detection provides?   How can we characterize how many models
predict similar binary parameter distributions?  Given a proposed model, how can we devise simple, phenomenological
diagnostics that characterize critical differences from neighboring models?

In this paper we describe a general yet practical method characterize model differences, using a coordinate-free
``generalized distance'' on the space of distributions.  Our discussion is  broadly applicable to experiments
where on the one hand experiments measure several properties of each event but on the other hand theory cannot produce comparable multivariate
distributions without significant computational cost.  Our method is statistically well-posed;  applies equally well to
coarse or fine sampling of the model space; can employ as many or few experimental measurables as desired; can be easily modified to
account for experimental error; and asymptotes to a maximally informative Fisher matrix analysis in the
limit of small statistical and systematic errors.  
  To illustrate this method, we employ a concrete toy model, with parameters qualitatively suitable for gravitational wave detection of a population
of merging compact binaries.   %
In Section  \ref{sec:KL} we review the theory underlying our diagnostic, the Kullback-Leibler (KL) divergence.    We
explain how, given a model, we can sort all other models by their ``proximity'' to it.  We describe the formal
relationships between our diagnostic, the log-likelihood,  and the Fisher matrix. 
Following other recent studies, we define an \emph{effective local dimension} to characterize the complexity of the
local model space. 
In Section \ref{sec:KLandObservations} we explore how the region of models consistent with observations decreases in
size as the number of observations increases.   
As an example,  in Section \ref{sec:ToyModel} we apply this method to a physically-motivated toy model for future
gravitational wave observations.

For simplicity, in this work we adopt a highly idealized model for gravitational wave detection, where our unchanging
detector has perfect success at identifying any binary inside
a sphere of radius $D_{max}$ and complete failure outside that distance.   We likewise assume mergers occur uniformly
in space and time.     More realistic models of time-dependent star formation and cosmologically significant reach will be
addressed future publications.

\subsection{Context and related work}

Though measured through other means, binary compact object merger rates are already used to constrain progenitor models, both
in the Milky Way \cite{psr-spinFb,PSmoreconstraints} and throughout the universe \cite{2010ApJ...722.1879M,PSgrbs-popsyn,grbs-short-KBROS-MaxNSMass,PSellipticals}. %
Though different populations may be selected by gravitational waves, notably binary black holes, qualitatively similar
methods can be applied to weakly constrain binary evolution using the first few detections \cite{2010CQGra..27k4007M}.   
Previous studies have also realized that the greater information available per detection allows even stronger constraints \cite{2010CQGra..27k4007M}.
Roughly speaking, the implied parameter distributions can be reconstructed nonparameterically, using the known
statistics of the detection process.  
For example, the binary pulsar merger rate is reconstructed with Poisson statistics \cite{Chunglee-nsns-1,psr-spinFb}.  
The binary parameter distribution can be similarly reconstructed, using either an ad-hoc or physically motivated kernel
to describe each point's ambiguity function \cite{2010PhRvD..81h4029M}.\footnote{In fact, parameter estimation methods
  applied to each detection provide a very detailed model for the ambiguity function associated with each event.}
We will not address the problem of distribution estimation; we take the recovered distributions and event rates as input.

Bayesian model selection theory has been extensively applied in gravitational wave physics, though rarely to
differentiate between progenitor population models.  
In the most comparable model, model selection was employed to assess how well a  discrete family of models for supermassive black hole
growth could be distinguished via their gravitational wave signal, as seen by pulsar timing arrays
\cite{2011PhRvD..83d4036S}.
In the context of gravitational wave astronomy, model selection has usually been applied to explore the significance of
proposed violations of General Relativity
\cite{2011PhRvD..83h2002D,2011PhRvD..84f2003C,gr-extensions-tests-Europeans2011}.   
Most directly comparable is a prototype study by  \citet{2010PhRvD..81h4029M}, which investigated how gravitational wave
events provide increasing evidence for  phenomenological mass distribution parameters. 
In astronomy, Bayesian model selection is applied far more extensively.  In one pertinent example, the known masses of
black holes and neutron stars can be fit to different models \cite{popsyn-constraint-StellarMassBHMassDistribution-Empirically-Farr2010,obs-ns-masses-Ozel2012}.

Other simpler methods have also been provided to differentiate between mass distributions and to determine which mass
distributions best fit the data.  For example, a simple one-dimensional Kolmogorov-Smirnov (KS) test can be applied to
compare observed binary (chirp) masses to each model's predicted distribution \cite{2004MNRAS.352.1372B,2003ApJ...589L..37B}.
Additionally, gravitational wave observations provide only some information.  Stellar binary populations and pulsar
recoil velocities also weakly constrain binary evolution \cite{popsyn-Tout-ConstraintsViaStellarPopulations2007,PSmoreconstraints}.    We will not address other comparison diagnostics  or multi-observable statistics
in this paper.

In principle, the gravitational wave signal encodes all multipole moments of the radiating spacetime.  Nonetheless, via
perturbation theory and separation of timescales \cite{ACST,gwastro-mergers-nr-Alignment-ROS-Methods,gwastro-mergers-nr-Alignment-ROS-IsJEnough}, the imprint of gravitational waves on ground-based detectors can
often be well-approximated by much simpler signals.    
Equivalently, the true high-dimensional signal manifold (a submanifold of $L^2$: all possible timeseries) is very close
to a much lower-dimensional submanifold.   
For weak signals, data analysis only requires the low-dimensional submanifold.  For strong signals, allowing fine
discrimination between signals, model estimation can require a higher effective dimension.   Both the  relevant effective dimension 
\cite{gwastro-mergers-HeeSuk-FisherMatrixWithAmplitudeCorrections} to a particular amplitude and the change in effective
dimension with amplitude \cite{gwastro-stochastic-coincidence-ROS-Stefanos} enter directly into gravitational wave data
analysis and parameter estimation.

\section{Diagnosing differences}
\label{sec:KL}

In this section we introduce a surprisingly simple but robust diagnostic that differentiates between predictions for the
number and nature of detections.   Our suggestions are motivated by gravitational wave data analysis, where severe
computational burdens\footnote{Severe computational burdens are associated both in generating astrophysical predictions
  and in calculating the selection biases that connect astrophysics to  parameter distributions for the observed
  sample.}  limit the ability to generate a large family of reliable predictions.  
Rather than a continuous model space that can be infinitesimally refined without effort, the model space we must work
with is invariably discrete and potentially poorly sampled.  We need tools that distinguish between models; that
identify potential undersampling of the model space; that estimate the number and nature of pertinent degrees of
freedom; and that do so in a well-understood  fashion that nonetheless can maximally exploit all available information.  
Moreover, these methods should ideally work without coordinates, owing to the extremely high dimension of the model
space; difficulty in a priori identifying natural variations; and computational intractability thorough sampling along
any single dimension.

Our diagnostic is surprisingly simple: the  log likelihood difference, averaged over all data realizations that one of
the two implies.  This difference factors into two contributions, measuring the difference in (a) the expected number and (b) the  ``shape''
(i.e., parameter distribution).  
Moreover, this diagnostic has close, provable connections to many statistical tools familiar from gravitational wave
data analysis, notably the likelihood and Fisher matrix.  
This diagnostic nonetheless allows us to identify ``similar'' systems blindly, even when the ``similarity'' region
extends over a large fraction of model space.   
In turn, the set of ``similar'' systems naturally defines characteristic shape \emph{variations}, letting us determine
the number and nature of pertinent degrees of freedom \emph{nonparametrically}. 

\subsection{KL divergence defined}
For our purposes, a binary evolution model makes two predictions: first, $\mu$, the average number of events our detector
should identify; and second $p(x)dx$, the probability distribution of different binary parameters among the set of detected events (e.g., $x$ consists of component masses,
spins, location, and orientation).\footnote{In practice, detector-dependent selection biases strongly modify the
  underlying parameter distribution.  For example, for gravitational wave data analysis, the detection range scales as
  the (chirp) mass to the $5/6$ power: high mass sources are vastly easier to find, though intrinsically rare.   For simplicity, we assume that both the detector and astrophysics are well-known
  and thus that $p(x)$ is completely determined.}  To be concrete, as each detection occurs independently of all others, the
probability that our detector will identify binaries with each $x_k$ between $x_k,x_k+dx_k$ is
\begin{eqnarray}
P(x_1\ldots x_n)d^n x &=&d^n x p(n|\mu) \prod_k p(x_k)  \\
p(n|\mu) &\equiv& \frac{\mu^n}{n!}e^{-\mu} 
\end{eqnarray}
Conversely, given a model $X=(\mu,p)$ and a set of events $d\equiv (x_1\ldots x_n)$, we can define a likelihood estimate $\hat{L}$
\begin{eqnarray}
\hat{L}(X) &\equiv& \hat{L}(X|d) =  \ln P(x_1\ldots x_n)
\end{eqnarray}
where for shorthand we omit explicit dependence on the data realization $d$.  
Modulo priors and model dimension penalties, models with higher peak $\hat{L}$ are more plausible estimates for the
generating process for $x_1\ldots x_k$ than models with lower $\hat{L}$.  
Suppose  each measurement is independently drawn instead from a fiducial model $X_*=(\mu_*,p_*)$.   Averaging over all measurements
implies
\begin{eqnarray}
\label{eq:Likelihood:Mean}
\left<\hat{L}(X)\right>_{X_*} &=&  \left<\ln p(\nObs |\mu) \right> + \left<\nObs\right> \int d^{\dparam} p_*(x) \ln q(x)
\end{eqnarray}
If the fiducial and test models $X,X_*$ are equal, the average $\left<\hat{L}\right>$ is the sum of (a) the entropy of
the poisson distribution plus (b) $\left<n\right> = \mu_*$ times the entropy of the parameter distribution $p$.  In the
more general case where $X\ne X_*$, the mean log likelihood $\left<\hat{L}\right>$ will be smaller than this bound.  We
characterize the decrease in expected log likelihood with the KL divergence.

For a general pair of probability distributions $p(x),q(x)$, the entropy $H_p$ and KL divergence $D_{KL}(p|q)$ are
defined by  \cite{2011RvMP...83..943V,stat-MacKay-BayesianInterpolation}
\begin{eqnarray}
H_p&=& -\int dx p \ln p \\
D_{KL}(p|q)&\equiv& \int dx p \ln p/q 
\end{eqnarray}
Roughly, the KL divergence characterizes the information gain the data must provide  to go from a prior $p$ to a
posterior $q$.  
The KL divergence is non-negative definite with $D=0$ if and only if $p=q$.  The KL divergence is \emph{not} symmetric.
Substituting into Eq. (\ref{eq:Likelihood:Mean}), the 
\begin{eqnarray}
\label{eq:Likelihood:Mean:ViaD}
\left<\hat{L}(X)\right>_{X_*}
 &=&  -[D_{KL}(\mu_*|\mu) +  H_{\mu_*}] 
 \nonumber \\ & & - \mu_*[D_{KL}(p_*|p)+H_{p_*}] 
\end{eqnarray}
where $D(\mu_*|\mu)$ is shorthand for the KL divergence between two poisson distributions:
\begin{eqnarray}
D_{KL}(\mu_*|\mu)&=& \sum_n p(n|\mu_*)[ \mu-\mu_*  + n\ln(\mu_*/\mu)] \nonumber \\
\label{eq:Dkl:Poisson}
&=& \mu-\mu_*  + \mu_*\ln(\mu_*/\mu) \\
H_\mu&=&- \sum_n p(n|\mu)[ -\mu + n \ln \mu - \ln n!] \nonumber \\
&\simeq& \frac{1}{2}\ln 2\pi e \mu \quad \mu \gg 1
\end{eqnarray}
In particular, given a fixed reference model with parameters $\lambda_*$, the expected difference in log likelihood between two candidate models
$\lambda_1,\lambda_2$ can be expressed as a sum of two contributions:
\begin{eqnarray}
-\delta L&=& -(\left<\hat{L}(X(\lambda_1))\right>_* - \left<\hat{L}(X(\lambda_2))\right>_*) \nonumber \\
&=& -(\left<\ln p(n|\mu_1)\right> - \left<\ln p(n|\mu_2)\right>) \\
&=&   [D_{KL}(\mu_*|\mu_1)-D_{KL}(\mu_*|\mu_2)] \nonumber \\ &&+\mu_*[D_{KL}(p_*|q_1)-D_{KL}(p_*|q_2)]  
\end{eqnarray}

The KL divergence therefore provides a simple, invariant diagnostic, quantifying differences between models.  Further,
the differences it identifies are statistically meaningful, connected to differences in (expected) log likelihood.
Finally, the differences \emph{factor}: the two terms tell us how to weight models' differences, on the one hand in rate
(the mean number of detections) and on the other hand in their predicted parameter distributions.

\subsection{Fisher matrix and local dimensionality}
One way to discriminate between models relies on  maximum likelihood.  Observations of viable models have $\hat{L}$
increase (with decreasing relative variance) as more observations $x_k$ accumulate.   For any given data realization, statistical fluctuations insure
that many models with only marginally smaller $\left<L\right>$ cannot be reasonably distinguished.
We therefore want to know how many models are ``nearby,'' in the sense that the average $\left<L\right>$ is within some
threshold of the value for the reference model itself.

For well-determined observations, the log likelihood has a narrow peak, defining a $d_{params}$-dimensional ellipsoid.
  In a small neighborhood surrounding the reference model $\lambda_*$,  the mean log likelihood can be expanded in series using
the Fisher matrix $\Gamma_{ab}$:
\begin{eqnarray}
\label{eq:def:Fisher:ViaL}
\left<L(\lambda)\right> \simeq \left<L(\lambda_*)\right> - \frac{1}{2} (\lambda-\lambda_*)_a(\lambda-\lambda_*)_b \Gamma_{ab}
\end{eqnarray}
where for brevity we use $(\lambda-\ldots)_a$ to denote the coordinate vector $(\lambda_a - \ldots_a)$ and similarly.  
In particular, given a threshold  $\Delta L$ in log likelihood, a coordinate volume of order $(\Delta
L)^{d_{params}/2}/\sqrt{|\Gamma|}$ has (median) log likelihood within $\Delta L$  of the maximum likelihood point.  
This coordinate volume can be very small and scale very favorably with $\Delta L$, given the many parameters $d_{params} \gtrsim 7$ commonly modified in binary evolution
\cite{StarTrack2,PSmoreconstraints,PSellipticals}.

In practice, however, many detections will be required to tightly confine all binary evolution model parameters.
Rather than an ellipsoid, a threshold $\Delta L$ simply restricts to some extended subspace.  Nonetheless, we can still
compute $\left<L\right>$ for any model and thus pair of models.   If we can sample the model space, then for each
reference model, we can still
quantify how many models ``look similar'': the coordinate volume $V(<\Delta L | \lambda_*)$ defined by 
\begin{eqnarray}
V(<\Delta L | \lambda_*) \equiv \int_{L(\lambda)-L(\lambda_*)< -\Delta L} d\lambda
\end{eqnarray}
For each model this relation defines a one-dimensional function $V(<\Delta L)$.     For very small $\Delta L$, the
parameter volume will scale as $V \propto (\Delta L)^{d_{params}/2}$ as described above.  At larger likelihood
differences, some parameters are weakly constrained while others are determined.  In this regime, we define an
``effective dimension'' $d_{eff}$ by 
\begin{eqnarray}
d_{eff} \equiv 2  \frac{d\ln V}{d\ln \Delta L}
\end{eqnarray}
On physical grounds, we anticipate $d_{eff}$ should usually decrease monotonically as $\Delta L$
increases.\footnote{The derivative of the volume is mathematically equivalent to the (logarithmic derivative of the)  density of states.  Many examples
  in condensed matter physics show the density of states and by implication $d_{eff}$ can behave unexpectedly in
  fine-tuned scenarios.}
For large likelihood differences, all models are consistent and $V$ converges to the whole parameter space.

In practice, the simulation space cannot be exhaustively explored: not enough computing power is available to sample all
possible likelihood differences.   Nonetheless, for each candidate reference simulation, the functions $V(\Delta L)$ can
be easily estimated by evaluating $\left<L\right>$ for all other simulations, building a histogram, and fitting accordingly.

To this point we have used a composite discriminant ($\left<L\right>$) involving both rate and shape.  One can also determine how many models
have a similar \emph{event rate alone} or \emph{parameter distribution alone} as our reference model.
To answer the first, let us translate a threshold on  $\Delta L_{rate}$, the contribution to the log likelihood from the
event rate alone, to an uncertainty in the event number $\mu$:
\begin{eqnarray}
\label{eq:Likelihood:Difference:RateAlone}
\Delta L_{rate} = - D_{KL}(\mu_*|\mu_1)  \simeq - \frac{1}{2} \mu_* [\ln(\mu/\mu_*)]^2
\end{eqnarray}
A threshold on $\Delta L_{rate}$ corresponds to a relative uncertainty $\delta \mu/\mu_* \simeq
\sqrt{2\Delta L_{rate}/\mu_*}$ in the event rate.    In other words, the event rate provides a single real measurable
parameter \footnote{Binary evolution comparisons are particularly simple when event rate is used as a
  coordinate.} ($\ln \mu$) which can be measured to an accuracy scaling as $\sqrt{\Delta L/\mu_*}$ (i.e., the poisson
limit $\propto 1/\sqrt{n}$,  if the likelihood threshold $\Delta L$ doesn't change with the number of events or location
on the parameter manifold).

By comparison with above and the process of elimination,  the shape diagnostic $D_{KL}(p_*|p)$ constrains  the
remaining $d_{eff}-1$ parameters.   Explicitly, the KL divergence $D_{KL}(p_*|p)$ between a reference distribution $p_*$ and perturbed
configuration $p$  can be expanded as
\begin{eqnarray}
D_{KL}(p_*|p)& =&  \int p_* \ln p_*/p 
 \nonumber \\
\label{eq:Dshape:CoordinateExpandNearPoint}
 &\simeq&   \frac{\delta \lambda_a \delta \lambda_a}{2}  \left< \frac{\partial^2}{\partial \lambda_a \partial \lambda_b} \ln p \right>
\end{eqnarray} 
in the limit of infinitesimal parameter change $\delta \lambda =\lambda-\lambda_*$ and assuming $p$ is never precisely zero.\footnote{A model where physics completely forbids a
  particular configuration can always be distinguished from one that does not, via a single observation with those
  conditions.  We will not discuss this limit here.} 
[The linear-order term cancels, as $\int p=1$ for all $\lambda$.]  In principle, the final expression  lets us
 calculate the contribution to the positive-definite  Fisher matrix $\Gamma_{ab}$ due to shape changes $\delta p$  in terms of tabulated shape distributions
$p(x|\lambda)$.  In practice, however,  numerical estimates of $p(x|\lambda)$ are rarely  accurate
enough to allow accurate second derivatives.   More critically, the parameter space has not been thoroughly enough
explored to permit this second derivative to be accurately evaluated, except in a handful of selected cases.

\subsection{Characteristic shapes versus tolerance}
Even the most complicated binary progenitor models have a finite number of parameters.  This number sets the number of
ways the distribution $p$ could change significantly.  For example, in the limit of arbitrarily small variations $\delta
\lambda$ about $\lambda_*$,  the change  $\delta p$ in $p$ can be computed via the $d_{params}$ independent logarithmic derivatives
$\partial_\lambda \ln p$.  
These variations prove particularly useful at identifying and quantifying the impact of each parameter,
and at identifying correlations.
For example, these variations determine how the entropy $H(p_*)$ and therefore likelihood of the reference model 
changes as we move $\lambda_*$ across the parameter space.
Conversely, if we fix the reference model $\lambda_*$, these (constrained) variations  determine the difference in log likelihood
between the reference model and our candidate point using Eq. (\ref{eq:Dshape:CoordinateExpandNearPoint}).
Finally, practically speaking these variations help us interpret the space of models and  determine what
measurements (i.e., what data)  best discriminate between them.
Some measure of the list of possible shape variations $\delta \ln p$ allowed near a particular reference configuration
$\lambda_*$ therefore proves exceptionally useful.  Unfortunately,  derivatives like $\partial_\lambda p$ and $\partial^2_\lambda p$ are too difficult to calculate
reliably when  standard Monte Carlo methods for progenitor model simulation are employed.  

We can nonetheless  estimate the space of allowed variations directly from a large but discrete sample of the parameter
space near some $\lambda$.   Specifically, given a reference model selected from a  large collection  of \emph{random,
  uniformly sampled} models, we first
eliminate all models with expected log likelihood farther than  $\Delta L$ (i.e., that lie outside $V(\Delta L|\lambda_*)$), assuming data ensembles are drawn from the
reference model.   The subset that satisfies this condition will randomly sample the consistent volume.  If the limiting
volume is sufficiently small as to be ellipsoidal, then the random sampling will ensure more samples along then
large eigendirections of $\Gamma$ and fewer in other directions.  We estimate (and rank) the plausible variations in the
neighborhood of $\lambda_*$ via a principal component analysis of the $N_{samp}$ distributions  $p_A =p_1\ldots p_{N_{samp}}$
where $A=1\ldots N_{samp}$ indexes the sampled points:
As described in more detail in the appendix, functional principal component analysis corresponds to the eigenfunction
problem for the correlation operator  \cite{mm-statistics-nonparametric-timeseries-RamsaySilverman}:
\begin{eqnarray}
\label{eq:def:Corr}
{\cal C}  &\equiv& \frac{1}{N_{samp}}\sum_{A=1}^{N_{samp}} \qmstate{\delta p_A} \qmstateKet{\delta p_A}
\end{eqnarray} 
where $\delta p_A(x) \equiv p_A(x)-p_*(x)$ and where we adopt a quantum-mechanics-motivated notation for brevity.  In
this notation, inner products are evaluated with flat norm
\begin{eqnarray}
\qmstateproduct{a}{b} \equiv \int dx a^*(x) b(x)
\end{eqnarray}
The most significant eigenfunctions and eigenvalues of  ${\cal C}$  tell us the most significant ways the predicted
parameter distribution $p$ varies in the sampled domain.  
In general this operation will admit many eigenvectors, potentially more than  $d_{params}$.  We expect the number of
marginally significant eigenfunctions to be particularly large when the constrained volume $V(\Delta L)$ is
significantly nonellipsoidal.

By construction, this operation will recover  $\partial_\lambda p$ as eigenfunctions in the limit of a
small ellipsoidal volume.    To prove this, expand each simulation's functional variation $\delta p_A$ in series about
$p_*$, using their known coordinate locations $\delta \lambda_{A,a}$:
\[
\delta p_A \simeq \delta \lambda_{A,a} \partial_{\lambda_a} p 
\]
Substituting this approximation  into the definition of ${\cal C}$, we find
\begin{eqnarray}
{\cal C} &\simeq& \qmstate{\partial_{\lambda _a} p}\qmstateKet{\partial_{\lambda_b} p} \frac{1}{N}\sum_A \delta
\lambda_{A,a}\delta\lambda_{A,b} \nonumber \\
&\simeq& \qmstate{\partial_{\lambda _a} p}\qmstateKet{\partial_{\lambda_b} p}  \Gamma_{ab}
\end{eqnarray}
In the second line, we are replacing the estimate of the coordinate correlation function by its expected value, known
from the shape of the ellipsoidal region.
The eigenfunctions of ${\cal C}$ are therefore the eigenvalues of $\Gamma$ dotted into derivatives $\partial_\lambda p$.

Generally, the principal components  naturally and invariantly characterize the degrees of freedom
available to us.
Critically, this method neither requires coordinates nor needs fine sampling: the principal components of a coarse grid
naturally characterize variation on the coarse scale.

\subsection{Effective dimensions greater than the number of model parameters?}
When the model space has indistinguishably small scales, the effective dimension can be significantly lower than the
number of parameters needed to specify the model.
At the other extreme, the effective dimension can also be \emph{higher}, either by the model manifold ``filling volume''
by folding on itself or by ``rotating'' through the infinitely many degrees of freedom possible in a space of
distributions.   To understand the latter, imagine the space of all possible parameter distributions $p(x)$ as  a
high-dimensional vector space;  the model space is some low-dimensional submanifold.  
  As we illustrate by example in Section \ref{sec:ToyModel}, that submanifold can ``rotate'' through many dimensions,
  with significant differences in the number of local degrees of freedom.
On the other hand, the model space (like a fractal or ergodic curve) may also wind back around near itself, creating
the appearance of greater complexity and higher dimension.   

In both scenarios, both approaches for calculating local dimension -- the effective dimension and a principal component
analysis -- will find model dimensions  \emph{greater} than the number of underlying parameters.  
The number of degrees of freedom identified by principal component analysis will be larger (possibly much
larger).  For example, as illustrated in Section \ref{sec:ToyModel}, principal component analysis treats distinct, nonoverlapping parameter distributions $p(x)$ as
two \emph{independent basis vectors}: all linear superpositions are allowed.

Whenever the effective dimension is large,  observations only weakly constrain our model space:  many if not all model parameters
cannot be constrained.  Instead, a purely phenomenological approach should be adopted, based on the range and
differences of distributions seen in simulations.
In special cases, a principal component analysis may suggest that some parameters can be constrained but other
degrees of freedom should be handled phenomenologically.  This situation can arise naturally when certain parameter
predictions are well-sampled and tightly connected to one or a few model parameters (e.g., the strength of supernova kicks on
pulsars; the masses of neutron stars) but the remainder are not.  We will address this scenario concretely in a
subsequent publication.

\section{Diagnostics and detection}
\label{sec:KLandObservations}

Knowledge of how much  $\hat{L}$ can fluctuate between data realizations helps determine what threshold to set for
likelihood differences $\Delta L$ for different $n$.  

For a specific experiment duration and scale, by construction  our diagnostic tells us how well a maximum-likelihood
statistic could distinguish between any two models.
A more sensitive or longer experiment will accumulate more events, increasing the average maximum likelihood,
reducing the relative likelihood range  $\Delta L/L$ that observations do not distinguish between, and therefore
shrinking the observatinally-consistent parameter volume.
How does the discrimination process scale with the number of events, on average?   How much do random fluctuations
impact the likelihood difference?  How does the size of the ambiguity region scale with the number of observations?

To address these questions requires nothing more than a model for how the average and standard deviation of $L$ change
with the expected number of events.  Specifically, if $X=(\mu,p)$ and $X_*$ are models, with $X_*$ the truth, we want to
know the statistical properties of the likelihood difference
\begin{eqnarray}
\hat{L}(X) - \hat{L}(X_*)
\end{eqnarray}
As each successive measurement is independent, the mean log likelihood difference between any two models grows linearly,
with a rate that depends on $X$ and $X_*$.  
 By contrast, fluctuations in the log likelihood difference can grow in complicated ways with the number of events, depending sensitively on how
 distinguishable the two internal parameter distributions are.    For example,  if the reference and test models are
 identical ($X=X_*$) the likelihood difference is trivially always zero, with zero variance.   By contrast, a if the
 reference and test models  have dramatically different parameter distributions (e.g.,
 $p/p_* \simeq 0$ in some region), the likelihood differences will generally be large and highly variable, depending on
 how often sample points occur in the region where the two models differ.  

Previous studies have described how to use either  the event rate or some parameter distribution (e.g., the mass
distribution) to differentate between models.  To connect to these previous studies, to simplify the calculations, and
to allow time to reflection on their implications, we
too explore how well partial information helps distinguish between progenitor scenarios.

\subsection{Average versus fluctuating likelihood 1: Event rate alone}
\label{sec:sub:PoissonAndLogEventRate}

Suppose our detector only identifies candidate events.  [For simplicity we continue to assume the detector operates with
  perfect confidence.]  
On average, each successive independent measurement changes the expected log likelihood $\hat{L}$ of a proposed model $X=(\mu,p)$ by a fixed amount:
\begin{eqnarray}
\left< \frac{\hat{L}(X) - \hat{L}(X_*)}{n} \right>  &=& \left< \frac{\ln p(n|\mu)/p(n|\mu_*)}{n} \right> 
\end{eqnarray}
Differences in likelihood between two models therefore grow linearly with (average) detected number (i.e., linearly with
time or range cubed).   As one would expect from Eqs. (\ref{eq:Likelihood:Mean}-\ref{eq:Dkl:Poisson}), to an excellent
approximation, the average change per event can be approximated by $D_{KL}(\mu_*|\mu)/\mu_* \simeq - (\ln
\mu/\mu_*)^2/2$ [see Eq. (\ref{eq:Likelihood:Difference:RateAlone})].

\ForInternalReference{
\sampleLinkWithLabelInline{math/ProbabilityAndMeasure/Probability/DiscreteStatistics/overview}{Sample problems:
  discrete statistics}
}
By contrast and as expected from standard Poisson statistics, fluctuations in the log likelihood scale as a smaller
power ($\sqrt{\mu_*}$) of the number of events.
To demonstrate this result in the notation provided below, the standard deviation of $\delta L(X) \equiv \hat{L}(X)-\hat{L}(X_*)$ about its mean value can be expressed in terms of
certain moments of the poisson distribution:
\begin{eqnarray}
\sigma^2_{\delta \hat{L}} &\equiv & \left< (\delta \hat{L})^2 - \left< \delta \hat{L}\right>^2 \right> \nonumber \\
 &=& \left< [\ln (p(n|\mu)/p(n|\mu_*))]^2 \right> - \left< \ln (p(n|\mu)/p(n|\mu_*)) \right>^2
\end{eqnarray}
The second term is precisely $D_{KL}(\mu_*|\mu)^2$ [Eq. (\ref{eq:Dkl:Poisson})]; the first term can be evaluated
directly:
\begin{eqnarray}
\left< [\ln (p(n|\mu)/p(n|\mu_*))]^2 \right> &=& \sum_n p(n|\mu_*)[ \mu-\mu_*  + n\ln(\mu_*/\mu)]^2 \nonumber \\
&=& (\mu-\mu_*)^2 + \mu_*(\mu_*+1) [\ln (\mu/\mu_*)]^2 \nonumber \\ 
 &+& 2 \mu_*(\mu-\mu_*) \ln (\mu/\mu_*) 
\end{eqnarray}
As a result, we conclude fluctuations in the likelihood difference between models are infinitesimal  when those models
make the same predictions, but can be significant when they make different predictions
\begin{eqnarray}
\label{eq:SigmaDeltaL:NumberOnly}
\sigma^2_{\delta \hat{L}} &=& \mu_* \left[ \ln\mu/\mu_* \right]^2  \simeq  2 D_{KL}(\mu_*|\mu)
\end{eqnarray}
Fluctuations in  $\delta L$ are smaller than its mean value whenever the number of detections are sufficiently large,
compared to the relative difference $\ln \mu/\mu_*$ between them:
\begin{eqnarray}
1> \frac{\sigma_{\delta \hat{L}}}{\left<\delta L\right>} &=& \frac{\sqrt{\mu_*} |\ln \mu/\mu_*|}{D_{KL}(\mu_*|\mu)}
\nonumber \\
 &\propto & \frac{1}{\sqrt{2 \mu_*} |\ln \mu/\mu_*| }  = \frac{1}{\sqrt{D_{KL}(\mu_*|\mu)}}
\end{eqnarray}

\noindent \emph{How do measurements improve with more points?}: By construction, measurements improve in proportion to
the rate at which the distinguishable volume decreases as the number of measurements improve.  Since the (average) log
likelihood difference increases linearly with the number of measurements, the ability of measurements to distinguish
between models is best characterized by the rate of increase in distinguishable volume with (average) likelihood: the effective dimension.

The local dimensionality of a model space is easily understood by
using $\ln \mu$ as a coordinate.  Suppose $g(\ln \mu) d\ln \mu$ is our model density (e.g., our prior probability on $\ln \mu$).  The
above criteron shows that by measuring $N$ events,  models can be distinguished to an accuracy of order $\ln \mu \simeq
1/\sqrt{2N}$.    If the prior probability in this small neighborhood is nearly constant, the local dimension is nearly
unity.  

By contrast, observations may support predictions at the upper or lower end of the range consistent with theory, where
$g(\ln \mu)$ is rapidly changing and dropping to zero.  In this region, observations have far greater discriminating
power.  In this case, the effective dimension reflects the functional form of the prior in near the sample point,
averaged over scales $\simeq 1/\sqrt{N}$ in $\ln \mu$.    

By way of concrete example, if $\ln \mu$ is a linear function of a high-dimensional bounded parameter space, then the
upper limit of $\ln \mu$ generally involves a either a ``corner''  or local extremum of $\ln \mu$.   As a result, the
prior density  $g$ will have a power-law decay near the cutoff, $g d\ln \mu\propto (\ln \mu/\mu_{max})^{\alpha-1} d\ln
\mu$.   If sufficiently many measurements isolate us to a small neighborhood of $\mu_{max}$, the effective dimension
will therefore be $2 \alpha$.\footnote{The effective dimension in general is twice the derivative of the logarithm of the
  cumulative distribution in a parameter against the logarithm of that parameter.  The factor of two arises when we
  interpret the cumulative distribution as a volume of a sphere and the parameter as a ``radius squared'' of the
  constrained volume.}

\subsection{Average versus fluctuating likelihood 2: Spatial dependence alone}
\label{sec:sub:DklAsQuantifyNumberNeededToDistinguish}
At the other extreme, two models that predict the same number $\mu=\mu_*$ of events on average can be distinguished if
they predict distinctly different parameter distributions.  
This seemingly degenerate case is surprisingly representative: detailed event rate calculations rely on highly uncertain
inputs, such as the total amount of star-forming gas and the fraction of stars forming in binaries, that impact all
predictions \emph{proportionally}.

In this limit, as before, each successive independent measurement changes the expected log likelihood of a proposed model $X$ by
a fixed amount, in this case by
\begin{eqnarray}
\left< \frac{\hat{L}(X) - \hat{L}(X_*)}{n} \right>  &=& \left< \ln p/p_* \right>  =D_{KL}(p_*|p)
\end{eqnarray}
As before, the log likelihood difference between the two models fluctuates between different data realizations.  Taking
care to distinguish between all the $n$ independent samples involved in the likelihood, we find
\begin{eqnarray}
\sigma_{\delta \hat{L}}^2 &=& \left< [\sum_k \ln p(x_k)/p_*(x_k)]^2 \right>  - \left<n \ln p/p_*\right>^2
\end{eqnarray}
The second term is simply $\mu_*^2 D_{KL}(p_*|p)^2$.  The first term can be reorganized into a sum over $n$ terms
quadratic in a function of the point $x_k$ and a sum over $n(n-1)$ terms where the two points are distinct:
\begin{eqnarray}
\left< [\sum_k \ln p(x_k)/p_*(x_k)]^2 \right> &=& \left<n(n-1)\right> \left<\ln p/p_*\right>^2 
\nonumber \\ &+& \left<n\right>\left< [\ln p/p_*]^2 \right> 
\end{eqnarray}
\begin{eqnarray}
&=&  \mu_*^2 [D_{KL}(p_*|p)]^2 + \mu_*  \left< [\ln p/p_*]^2 \right>  \nonumber \\
\end{eqnarray}
Fluctuations in $\delta \hat{L}$ are smaller than its mean value whenever the number of detections is sufficiently large
compared to a ratio characterizing how different the two parameter distributions $p,p_*$ are:\footnote{In the right hand side, both the numerator and the denominator are zero if and only if $p=p_*$.}
\begin{eqnarray}
\label{eq:SigmaDeltaL:ShapeOnly}
1> \frac{\sigma_{\delta \hat{L}}}{\left<\delta L\right>} &=& \frac{ \left< [\ln p/p_*]^2 \right>^{1/2}  }{- \sqrt{\mu_*}D_{KL}(p_*|p)}
\end{eqnarray}

Both $\left< [\ln   p/p_*]^2 \right> $ and $\left< \ln p/p_* \right> = - D_{KL}(p_*|p)$ characterize differences in two
distributions.  In general, the two do not agree, as the first therefore is inevitably larger than the
square of the second:
\begin{eqnarray}
\left< [\ln   p/p_*]^2 \right> & =& \left<\ln p/p_*\right>^2 +  \left< [\ln   p/p_* - \left<\ln p/p_*\right>]^2 \right> 
\end{eqnarray}
For example, substituting $p,p_*$ both normal distributions with zero mean and standard deviation $\sigma,\sigma_*$, we
can easily show
\begin{eqnarray*}
\left< [\ln   p/p_*]^2 \right> = \left< [\ln   p/p_*] \right>^2 + \frac{ (\sigma_*^2-\sigma^2)^2}{2\sigma_*^4}
\end{eqnarray*}
Critically, in the limit $\sigma\rightarrow \sigma_*$, the
second term scales linearly as $D_{KL}(p_*|p)$, not quadratically.  We anticipate similar behavior in general.
Specifically, we can always choose  $q_p\equiv - D_{KL}(p_*|p)$ as one coordinate for the distribution space $p_*$.  Since both
terms  go to zero (smoothly), both can be expanded in Taylor series, in integer powers $q_p^s$ for
$s=1,2,\ldots$  of $q_p$.  
As in the event-rate-only case [Eq. (\ref{eq:SigmaDeltaL:NumberOnly})], a leading-order linear term $\propto
D_{KL}(p_*|p)$ is expected in general.

As in the rate-only case [Eq. (\ref{eq:SigmaDeltaL:NumberOnly})], the linear-order term dominatees the ratio on the
right side of  Eq. (\ref{eq:SigmaDeltaL:ShapeOnly}) in the neighborhood of small $D_{KL}(p_*|p)$.  Expanding the ratio in
series, we find
\begin{eqnarray}
 \left<[\ln p/p_*]^2\right> &\equiv& c_o D_{KL} + c_1 D_{KL}^2 + \ldots \\
\frac{\sigma_{\delta \hat{L}}}{\left<-\delta L\right>} &\simeq& \frac{1}{\sqrt{\mu_*}}[  \frac{c_o}{\sqrt{D_{KL}(p_*|p)}}
 + \frac{c_1 D_{KL}(p_*|p)}{2 \sqrt{c_o}} + \ldots ]
\end{eqnarray}

This formal expression has an intuitive interpretation: to distinguish two very similar parameter distributions $p,p_*$
(i.e., two distributions with $D_{KL}(p_*|p)\simeq 0$) requires extracting many sample points from the distribution
($\mu \gg 1/D_{KL}$).  The more similar the two distributions, the more samples are required.

\noindent \emph{How do measurements improve with more points?}: Like the likelihood, the KL divergence $D_{KL}(p_*|p)$ has a
local extremum.  Thus, as with the likelihood, if the collection of indistinguishable models is a small ellipsoid,  the parameter volume inside a contour of constant $D_{KL}$ scales as
$D_{KL}^{d/2}$ for $d$ the total parameter space dimension.   
In particular, if we choose our contour of constant $D_{KL}$ to satisfy Eq. (\ref{eq:SigmaDeltaL:ShapeOnly}),  then the
parameter volume will scale as 
\[
V \propto 1/\mu_*^{d/2} \; .
\]

By contrast, if a wide range of models are indistinguishable, then  the distinguishable volume $V$ must scale as a
smaller power $V\propto \mu_*^{-d_{eff}/2}$ for $d_{eff}<d$.

\subsection{Average versus fluctuating likelihood 3: Models with different rates and numbers}
In general, models predict both different numbers and distributions of events.
As above, on average, each successive independent measurement changes the expected log likelihood $\hat{L}$ of a proposed model $X=(\mu,p)$ by a fixed amount:
\begin{eqnarray}
\left< \hat{L}(X)/n \right>  &=& \left< \frac{\ln p(n|\mu)}{n} \right>  - [H_{p_*} + D_{KL}(p_*|p)]
\end{eqnarray}
Differences in likelihood between two models therefore grow linearly with (average) detected number (i.e., linearly with
time or range cubed).

\begin{widetext}
By contrast, fluctuations in log likelihood do not scale as a simple power of the expected number of events ($\mu_*$).  
As an example, consider the statistics of a single likelihood, assuming data is drawn from a reference model $X_*=(\mu_*,p_*)$:
\begin{eqnarray}
\hat{L}(X) &\equiv& \ln p(n|\mu)  + \sum_{k=1}^n \ln p(x_k) 
\end{eqnarray}
The standard deviation of $\delta L(X) \equiv \hat{L}(X)-\hat{L}(X_*)$ about its mean value can be directly evaluated.
After some algebra, we find 
\begin{eqnarray}
\sigma_{\delta \hat{L}}^2 &\equiv& \left<(\delta \hat{L} - \left<\delta\hat{L}\right>)^2 \right>  = \left< \delta \hat{L}^2 \right> -
\left<\hat{L}\right>^2 \nonumber\\
&=& \left<(\ln p(n|\mu)/p(n|\mu_*))^2 \right> - \left<\ln p(n|\mu)/\ln p(n|\mu_*)\right>^2 
  + [2 \left<(n-\mu_*)\ln p(n|\mu)/p(n|\mu_*)\right> ] \left<\ln p/p_*\right>
  +  \mu_* \left<(\ln p/p_*)^2 \right> \nonumber \\
&=& \mu_* \left[ |\ln \mu/\mu_*|^2 +  \left<(\ln p/p_*)^2 \right> + 2 (\ln \mu/\mu_*) \left<\ln p/p_*\right> \right]
  \nonumber \\
&=& \mu_*\left[ (\ln \mu/\mu_*  + \left<\ln p/p_*\right>)^2 
+ 
  \left<(\ln p/p_*  - \left<\ln p/p_*\right>)^2 \right>  
 \right]
\end{eqnarray}
\end{widetext}
where in the final line we regroup terms into a sum of two manifestly positive-definite quantities.  
In the second to last expression, the first two terms have been described previously and characterize fluctuations in $\delta \hat{L}$
when the two models predict different event rates or parameter distributions, respectively.  The new third term simply
reflects necessary correlations, implied by the fact that more samples will necessarily reveal more shape information on
average.   
Critically, all three terms scale in proportion to $\mu_*$, when the relative event rates ($\mu/\mu_*$) are
fixed.

\begin{figure}
\includegraphics[width=\columnwidth]{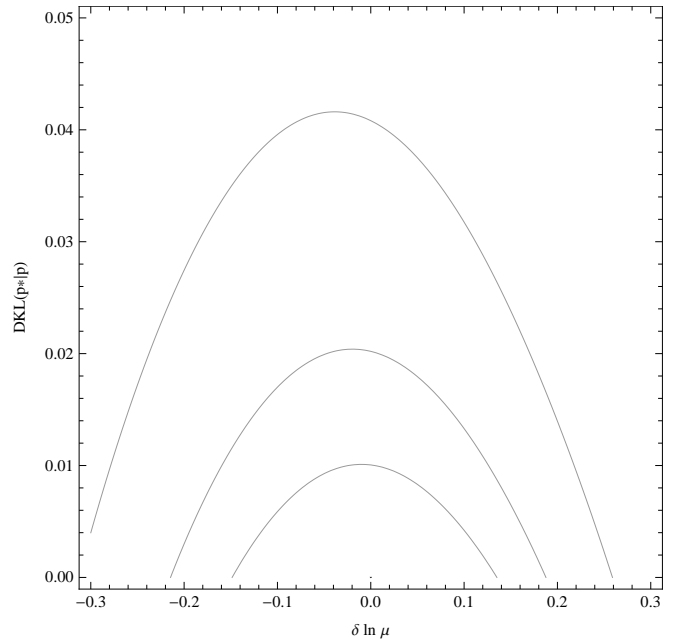}
\caption{\label{fig:PrototypeConstraint:Gaussian:1dFixSigma}\textbf{Constraint contour}: Contours of the ratio $\sigma_{\delta L}/\left<\delta L\right>$ versus $q_p
  \equiv D_{KL}(p_*|p)$ and $q_\mu \equiv \ln \mu/\mu_*$ for a simple two parameter model, with an arbitrary event rate
  $\mu$ (shown here for $\mu=100$) and gaussian parameter distribution $p$ of unit  standard deviation and arbitrary mean.  To a good
  approximation, the ratio $(<\delta L>/\sigma_{\delta L})^2 $ is  $\mu_*\times (q_\mu^2/2 +
q_p)^2/((q_\mu-q_p)^2+2 q_p) $.
Contours shown correspond to
  $\sigma/\left<\delta L\right> = 1/\sqrt{2},1,\sqrt{2}$, going from outermost inward.
Models with more degrees of freedom have constraint contours in a higher-dimensional submanifold, involving more
parameters than just the simple ``shape and rate'' diagnostics $q_\mu,q_p$.
}
\end{figure}

Precisely as before,  the condition $1\simeq \sigma_{\delta L}/\left<\delta L\right>$ defines a
surface in model space, allowing us to determine how many measurements are needed to distinguish two models.   Using $q_\mu = \ln
(\mu/\mu_*) $ and $q_p=D_{KL}(p_*|p)\ge 0$ as local coordinates, this expression  identifies a mishshapen box about the
origin in the
$q_\mu,q_p$ half-plane.  
Evidently, in the limit $q_\mu \ll q_p$ -- for models that predict similar event rate -- the constraint reduces to
Eq. (\ref{eq:SigmaDeltaL:NumberOnly}).
Conversely,  in the limit $q_p \ll q_\mu$ -- for models that predict similar parameter distributions -- the constraint
reduces to Eq.  (\ref{eq:SigmaDeltaL:ShapeOnly}).
Figure  \ref{fig:PrototypeConstraint:Gaussian:1dFixSigma} shows an example of the constraint surface, for a family $p$ of gaussian
distributions with fixed variance and arbitrary mean.

Whether calculated crudely (as a rectangle in $q_\mu,q_p$) or more precisely, the parameter volume inside the contour
scales as the local dimension.  If the constraint contour is sufficiently small to be ellipsoidal, then the constraint
volume must scale as 
\begin{eqnarray}
 V \simeq 1/\mu^{d/2}
\end{eqnarray}
where $d$ is  the total number of parameters in the model (e.g., all shape parameters plus the event rate parameter
$q_\mu$).
For weak constraints, the constraint volume will scale according to  some smaller effective dimension.

\hidetosubmit{
\subsection{Fluctuations and confidence intervals}
* If we just have to worry about fluctuations in number  \editremark{XX}
\begin{eqnarray}
\sigma_{\delta L}^2&=& \left< (L- \bar{L})^2\right> =  \mu_* [\ln (\mu/\mu_*)]^2
\end{eqnarray}
}

\hidetosubmit{

\subsection{Evidence and thermodynamic integration (*)}

* Key point: how confident can we be in a model?  Usual idea is to integrate out the full probability over some model
subspace, calculate ``evidence''

* KL divergence provides a natural way of doing so (it's automatically related).  Also, right now not talking about data
realizations, so EVIDENCE ISN''T RELEVANT YET

* But do we DO anything with it?  Not really.

(2) \editremark{relate to evidence: evidence as one partition function: idea of using generating function, with evidence as being one particular derivative}
}

\section{Applying constraints to a toy model for binary formation}
\label{sec:ToyModel}

Up to now, our discussion applies to any data modeling problem for poisson-distributed events with costly theoretical predictions.  In this
section we will develop a concrete example, using distributions, scales, and parameters motivated by gravitational wave
data analysis.   To remove any ambiguity or approximation, we choose  analytically tractable parameters (i.e., one-dimensional gaussian
chirp mass distributions).  
Using this example, we illustrate the ambiguity region versus likelihood threshold, emphasizing  how that region changes shape and
effective  dimension.   We explain how a small number of events constrain only one dimension; how many events constrain
all dimensions; and, depending on parameters, an intermediate sample size can constrain different things.   
We also  use of principal component analysis to identify and rank relevant mass distribution changes.  In this
case,  the effective dimensionality identified with these two techniques (volume scaling and principal components) is comparable.

\subsection{Model family}

The next generation of gravitational wave detectors should soon recover several tens to hundreds of events per year
\cite{LIGO-Inspiral-Rates,PSellipticals,popsyn-LowMetallicityImpact-Chris2008,popsyn-LowMetallicityImpact-Chris2010,popsyn-LowMetallicityImpact2-StarTrackRevised-2012}.
Subsequent detector generations with cosmologically significant reach should recover hundreds of thousands of events per
year  \cite{2011CQGra..28i4013H,2010CQGra..27s4002P,gw-detectors-ET-ScienceDocument}.
Given several orders of magnitude range, we treat the expected sensitivity of the detector as a tunable parameter and
discuss scenarios with $10, 10^2,10^4$ candidate events.

Gravitational wave detectors are exceptionally sensitive to massive binaries formed throughout the unvierse
\cite{LIGO-Inspiral-Rates,PSellipticals}, including exceptionally rare but massive black holes formed in low metallicity
environments \cite{popsyn-LowMetallicityImpact-Chris2008,popsyn-LowMetallicityImpact-Chris2010,popsyn-LowMetallicityImpact2-StarTrackRevised-2012}.
If binary black holes are included, theory currently only weakly constrains the space of plausible detected mass
distributions.\footnote{Observations of galactic black hole binaries can constrain the number and nature of low
  mass black hole binaries; see \citet{popsyn-constraint-StellarMassBHMassDistribution-Empirically-Farr2010}.}
Observations have far more tightly constrained neutron star masses and the binary neutron star mass distribution
\cite{obs-ns-masses-Ozel2012}.  
Motivated by that sample, for simplicity we adopt a single  gaussian (chirp) mass distribution centered on a
preferred peak value.\footnote{The chirp mass is ${\cal M}_c = (m_1 m_2)^{3/5}/(m_1+m_2)^{1/5}$.   For a canonical $1.4M_\odot+1.4
  M_\odot$ double neutron star binary, the chirp mass is $\simeq 1.2 M_\odot$. }
For simplicity and owing to detector limitations, we limit our discussion to a single mass measurement per binary.
While gravitational wave detectors in principle constrain both component masses and spins, strong correlations exist
between most recovered parameters \cite{1995PhRvD..52..848P}.  By contrast, even allowing for these correlations, the
chirp mass of a neutron star binary is invariably recovered with an accuracy much smaller than the mass
distribution's width.

We therefore examine  the following model space, parameterized by $\lambda = \{l_R,\bar{x},\sigma\}$ and a common
parameter $\bar{\mu}_*$:
\begin{eqnarray}
\mu &=& e^{lR} \bar{\mu}_* \\
p(x) &=& \frac{1}{\sqrt{2\pi \sigma}} e^{-(x-\bar{x})^2/2\sigma^2}
\end{eqnarray}
where the common factor $\bar{\mu}_*$ controls the overall experiment sensitivity (e.g., duration and range).   The KL
divergence between two one-dimensional gaussians has the simple form
\begin{eqnarray}
D_{KL}(p_*|p) &=& \int p_* \ln p_*/p  \nonumber \\
 &=& \ln \frac{\sigma}{\sigma_*}   
- \frac{ \left< (x-\bar{x}_*)^2 \right>}{2\sigma_*^2}
+ \frac{ \left< (x-\bar{x})^2 \right>}{2\sigma^2} \nonumber \\
&=& \ln \frac{\sigma}{\sigma_*}   
- \frac{1}{2}
+ \frac{ (\bar{x}-\bar{x}_*)^2 + \sigma_*^2}{2\sigma^2} 
\end{eqnarray}
Combining this expression and Eq. (\ref{eq:Likelihood:Mean:ViaD}), we can express the average log likelihood difference between the reference
model and any proposed model as 
\begin{eqnarray}
\label{eq:GaussianModel:AvDiffL}
\left<L(X) - L(X_*) \right> &=&  - D_{KL}(\mu_*|\mu) - \mu_* D_{KL}(p_*|p) \\
&=& -\bar{\mu}_* \left[
  e^{l_R}   - l_R  + \ln \sigma + \frac{\bar{x}^2+1}{2\sigma^2}- \frac{3}{2}
\right] \nonumber 
\end{eqnarray}
where without loss of generality we adopt $(\bar{x}_*,\sigma_*)=(0,1)$.
Finally, after some algebra, the standard deviation of the log likelihood about this mean value is
\begin{eqnarray}
\frac{\sigma_{\delta L}}{\mu_*} &=&
 \frac{3 + 6 \bar{x}^2 + \bar{x}^4}{4 \sigma^2}  + \frac{3+4 l_R(1+l_R) - 4(1+2y-\ln \sigma)\ln \sigma}{4} 
\nonumber \\
 &-&  \frac{1}{2\sigma^2}[3 + \bar{x}^2 + 2(1+\bar{x}^2)l_R -2 (1+\bar{x}^2)\ln \sigma ]
\end{eqnarray}

As we have seen previously, the interpretation of model constraints is simpler in coordinates
$q_\mu \equiv \ln \mu/\mu_*$ and $q_p \equiv D_{KL}(p_*|p)$ that characterize how strongly the event rate ($q_\mu$) and
parameter distribution ($q_p$) differ from the true value.   By eliminating $\bar{x}$ and $\mu$, we can explicitly
re-express both $\left<\delta L \right> $ and $\sigma_{\delta L}$ in terms of $q_p,q_\mu$ and $\sigma$:
\begin{eqnarray}
- \left<L\right> &=& \mu_* [q_p + e^{l_R}-(1+l_R)] \\
\sigma_{\delta L}^2 &=&  \frac{\mu_*}{2} [ 1+ 2 (l_R-q_p)^2  
 \nonumber \\  && - \sigma^{-4} + 4\sigma^{-2}(1+\ln \sigma)]
\end{eqnarray}
  Not all parameter combinations are realizable: $q_p$ must not only be positive, but it must be
greater than
\begin{eqnarray}
q_p \gtrsim Q(\sigma) \equiv \ln \sigma + \frac{1-\sigma^2}{2\sigma^2}
\end{eqnarray}

\begin{figure}
\includegraphics[width=\columnwidth]{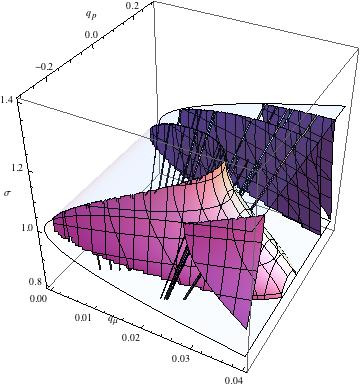}
\caption{\label{fig:PrototypeConstraint:Gaussian:3d}\textbf{Constraint contour (3d)}: For the three-parameter detection
  model described in the text (event rate, average mass $\bar{x}$, and mass distribution width $sigma$), contours of
  $\left<\delta L\right>/\sigma_{\delta L}$ versus $q_\mu$ (quantifying how different the event rate is from injected),
  $q_p$ (quantifying the difference in shape) and $\sigma$.
 In these coordinates, $q_p$ must be greater than a certain function of $\sigma$, indicated by the transparent contour
 and described in the text.
}
\end{figure}

Figures \ref{fig:PrototypeConstraint:Gaussian:1dFixSigma} and \ref{fig:PrototypeConstraint:Gaussian:1dFixSigma:WithBar} show contours of the ratio $\left<L\right>/\sigma_{\delta L}$ versus just $q_\mu$ and $q_p$, for the
special case $\sigma=1$.  The contours isolate a small region around the origin.   
More generally, Figure \ref{fig:PrototypeConstraint:Gaussian:3d} shows contours of the same function, allowing all three parameters to vary.  In this
more general case, as indicated by the transparent contour in this figure, the requirement that $q_p$ be greater than a
function of $\sigma$ excludes everything below a parabolic cut through the $q_p,\sigma$ plane.

\subsection{Example 1: Unknown median mass and moderately known event rate}

\begin{figure}
\includegraphics[width=\columnwidth]{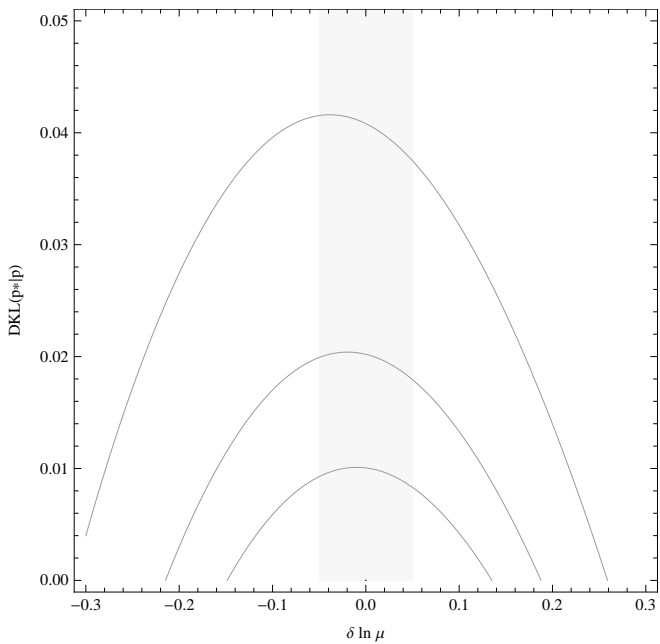}
\caption{\label{fig:PrototypeConstraint:Gaussian:1dFixSigma:WithBar}\textbf{Why the effective dimension increases}: As
  Figure \ref{fig:PrototypeConstraint:Gaussian:1dFixSigma}, except we overlay a band suggesting some a prior
  prior, here on event rate.  
}
\end{figure}

By combining this elementary model with different priors, we can illustrate the general properties described earlier in
the text.   As a highly unrealistic but illustrative example,  let us assume a high degree of prior information about
the event rate, illustrated by the gray band in Figure \ref{fig:PrototypeConstraint:Gaussian:1dFixSigma:WithBar}, and
complete knowledge of the mass distribution \emph{width} $\sigma$, but no information about the median value $\bar{x}$.
In this case, the first few detections will primarily inform us about the previously completely unknown value
$\bar{x}$.  Early on, the model space has \emph{one} effective dimension.  With only a handful of detections available, we can
effectively treat the event rate as known.  Graphically in Figure
\ref{fig:PrototypeConstraint:Gaussian:1dFixSigma:WithBar}, the  prior region defines a surface much narrower than
or constraint contours.   As those contours shrink with increasing numbers of events, we gain information only in
reduced ``vertical'' extent (via a smaller range allowed for $q_p$)

Eventually, once the number of detections $N$ is large enough to pin down the rate to better than the prior range
$\Delta \ln \mu$ (i.e., $N \gtrsim (\Delta\ln \mu)^{-2}$),  successive observations provide information about both the
mass distribution and the event rate.
Successive observations now fully constrain \emph{all two} dimensions of this model.

The transition from an effectively one-dimensional to two-dimensional problem is easily identified from the number of
models with different likelihoods, consistent with the prior.   For large likelihood differences, the number of models scales as $V \propto (\Delta
L)^{1/2}$, where by ``number'' I mean the volume relative to $(x,\mu)$ space in uniform measure.  By contrast,  for
small likelihood differences, the number of models scales as $V \propto (\Delta L)$.
Finally,  in the neighborhood of the optimum point, a principal component analysis also identifies the dominant
shape (the normal distribution $p$) and shape variation (a function $\propto \partial_x p$, corresponding to a shift in local maximum). 

To demonstrate this method unambiguously, we have  performed a Monte Carlo study, where we explicitly generate
synthetic gaussian realizations drawn from our model, estimate $\hat{L}$ for a large collection of candidate parameters,
 sort models by log likelihood, and use the ad-hoc $\Delta L$ threshold proposed earlier to estimate confidence
 intervals.  
The results of this toy model study behave precisely as expected and described above. 

\subsection{Example 2: Weak three-dimensional prior}

For binary neutron stars, by contrast, the event rate is very weakly constrained; by contrast, observations of pulsars
and X-ray binaries in the Milky Way relatively tightly constraint the mass distribution (mean and variance) for merging
binary neutron stars \cite{obs-ns-masses-Ozel2012,2007PhR...442..109L}.
Taking the tight neutron star mass distribution at face value,  we can use Figure
\ref{fig:PrototypeConstraint:Gaussian:3d} to determine the point at which gravitational wave measurements will provide
new information about the mass distribution.  For example, assuming the (relative) variance $\sigma$ is known to $10\%$ and $\bar{x}$ to $5\%$,
our prior volume in \emph{shape} parameters is small:  $q_p$ less than $ 10^{-3}$ and  $\sigma\simeq
[0.9,1.1]$.  
In this case, the first gravitational wave measurements will provide information only about the event rate, until at least
$N\simeq 100$ events  are observed; see  Figure \ref{fig:PrototypeConstraint:Gaussian:1dFixSigma}, which is scaled to
precisely $100$ events.
Of course, this condition could equally well and far more easily be derived from an understanding of gaussian
statistics.  Our methods, however, generalize to arbitrary distributions.

Alternatively, the (chirp) mass distribution for binary neutron stars could be substantially wider and biased \cite{popsyn-constraint-StellarMassBHMassDistribution-Empirically-Farr2010,obs-ns-masses-Ozel2012}: the
relevant supernova mechanisms could differ, accretion (slightly) changes each compact object's mass, et
cetera \cite{popsyn-LowMetallicityImpact2-StarTrackRevised-2012,StarTrack2}.  
More conservatively, one can adopt a mass distribution with median known to only $20\%$ and variance known to $50\%$. 
Again, either using gaussian statistics or Figure \ref{fig:PrototypeConstraint:Gaussian:3d}, one can deduce that far
fewer events are needed before gravitational wave observations will constrain the neutron star mass distribution.

\hidetosubmit{
\subsection{Dimension and characteristic variations}

\editremark{expected scaling}: this is a 3-parameter model.  A priori we can expect physical cutoffs int he range of
$\sigma$ and particular $\bar{x}$ (average mass). Taking that into account the model's effective dimension can quickly
drop well below unity.

\editremark{Consider}

1) 
* By implication, plot of effective dimension vs $\Delta L$ (by explicit calculation of coordinate volume)
Need to put edges on

*dimension: explain how effective dimension depends on the problem: 

** if similar mean and variance, at low rates the effective d should be nearly 1 (i.e., rate)

** as we get more events (i.e., as $\mu_*$ increases), we can probe multiple parameters

* EDGE EFFECTS: if we are near the edge (in general, near some concrete thing cutoff) we can be MUCH MORE INFORMATIVE,
because there are fewer models consistent with us

** demonstrate that by putting in a hard edge on the mass distribution.

* PCA-ing: what the PCA eigenvalues look like, for different thresholds.
}

\section{Implications for binary evolution and gravitational wave astronomy}

Using the above analysis as a guide, we can anticipate how future gravitational wave detection events will inform our
understanding of binary evolution.

\subsection{Tight constraints and rapidly improving performance}
As parameter distributions potentially encode \emph{infinitely} many degrees of freedom, these distributions can
completely encode all the details of formation scenarios.    Of course, our ability to directly constrain the
event rate and \emph{distribution} (nonparametrically) is highly limited by the number of events.  Realistic model
spaces are far smaller, however, allowing us to establish extremely tight constraints with a relative handful of
events.  The
fraction of models consistent with the data decreases as a high power of the number $n$ of measurement events -- conceivably, as
fast as $1/n^{d/2}$ (i.e., each parameter is independently constrained to an accuracy $1/\sqrt{n}$).  This exponent
reflects the (local) complexity of the model space, averaged over the set of predictions that cannot be distinguished
from our data.   As observations discriminate between finer details, the exponent increases, as the volume of
indistinguishable models grows smaller.

Unfortunately, the degree and rate of improvement depends strongly on what nature provides.  To use a simple example,
assume only the number of events can be measured, then compared to some single-peaked distribution like the ones
provided in \citet{PSellipticals}.   Observations that support the most plausible value are least informative: most
predictions cluster  there.   Subsequent observations  improve our certainty in the event rate by
$1/\sqrt{n}$ and reduce the fraction of consistent models by the same proportion.  
At the other extreme, if observations support an event rate near the limits of what our models predict, only a tiny
fraction of models can be consistent.  A range of event rate exist where further observations will reduce the fraction of models consistent
with observations faster than $1/\sqrt{n}$.  
To enable the tighest constraints, event rates and mass distributions should depend sensitively on several model parameters.
Previous studies suggest  both the number of events and their masses depend sensitively on assumptions \cite{PSconstraints,popsyn-LowMetallicityImpact2-StarTrackRevised-2012}.

\subsection{Rate and shape as two coordinates}
In this paper we propose adopting two specific coordinates  in the neighborhood of any
point in a proposed compact binary progenitor model parameter space: a ``log rate'' coordinate $\ln \mu$ and a ``change
in distribution shape'' coordinate $D_{KL}(p_*|p)$.    We recommend these coordinates because they are both calculable
and in direct relation to an intuitively obvious coordinate: the expected log likelihood difference between a model and
the truth [Eqs. (\ref{eq:Likelihood:Mean:ViaD}) or (\ref{eq:GaussianModel:AvDiffL})].   The full model space can be
parameterized with these two and any remaining $d-2$ coordinates.  
Qualitatively, both   $D_{KL}(p_*|p)$ or $\left<L\right>$ define isocontours that surround the local extremum: they are
the quantity being extremized and thus more like a ``separation squared'' than a coordinate free to take on any value.

Our coordinates quantify the effort needed to distinguish rate and shape.   
The KL divergence $D_{KJL}(p_*|p)$ quantifies on average how many measurements $\mu_*$ are required to distinguish two similar
distributions ($\mu_* \simeq 1/D_{KL}(p_*|p)$; see Section \ref{sec:sub:DklAsQuantifyNumberNeededToDistinguish}).
On the other hand, Poisson errors naturally limit how reliably we can measure the log of the event rate ($\delta \ln \mu
\propto 1/\sqrt{\mu_*}$; see Section \ref{sec:sub:PoissonAndLogEventRate}).  
Finally,  these coordinates let us pheneomenologically identify degrees of freedom.

While in practice we recommend these coordinates to phenomenologically identify relevant degrees of freedom, for conceptual purposes we strongly recommend one think in terms of
some underlying parameters $\lambda$ and, as necessary, a Fisher matrix to relate them to expected log likelihoods and
$D_{KL}$ [Eq. (\ref{eq:def:Fisher:ViaL})].  
Why?  Conceptually, the $D_{KL}$ coordinate is an (expected) log likelihood difference between the best-fit model and a
candidate.  Assuming we parameterize our model space with $\left<L\right>$ and $d-1$ other coordiantes, each choice of
$d-1$ coordinates defines some submanifold, which has a point of ``closest approach'' (i.e., largest likelihood) to the (unique) best-fitting model.
In other words, for each combination of the non-likelihood coordinates, a maximum value of $\left<L\right>$ exists,
leading to an excluded region in the coordinate space mirroring the local extremum, as in Figure \ref{fig:PrototypeConstraint:Gaussian:3d}.
Equivalently, this choice of coordinates has a singular Jacobian at the local extremum: $D_{KL}\simeq 0$ smoothly there
[Eq. (\ref{eq:Dshape:CoordinateExpandNearPoint})].   
For this reason, the coordinate $D$ cannot be used when counting dimensions for the volume scaling arguments (i.e.,
$V\propto \mu_*^{-d/2}$) unless the Jacobian is  taken into account and suitably integrated.

We anticipate models will be easiest to distinguish when we make full use of all available information.  In the language
of the previous sections, the KL divergence $D_{KL}$  between the predictions of two sets of model parameters $\lambda$ will increase when as many predictions as possible are included
in the measurement space $x$.

\subsection{Practical monte carlo to differentiate between models}
To this point our analysis has bene purely theoretical.  We will provide a more concrete analysis in a subsequent
publication.  However, in a previous study  \cite{2010CQGra..27k4007M}, we
have already used this method to quantify how often different members of a small model population could be
distinguished; see, e.g., their Figure 2.  In that study, synthetic data was generated according to each compact binary
model.  Each synthetic data set was compared with all its neighbors; if the likelihood difference was large enough
compared to the expected magnitude of fluctuations, they were assumed distinguishable.

Based on our discussion above, this study's results make perfect sense.   
First and foremost, because different models predicted different numbers of events, models that predicted many events
had only a handful of neighbors, at best, that could never be distinguished from them.  Moreover, as the predicted
number of events increased, the fraction of distinguishable models decreased, as expected (i.e., the fraction should
scale as $1/N^{d_{eff}/2}$ for $d_{eff}$ some effective dimension).
Unfortunately, the local error ellipsoid and therefore effective dimension depend on the reference point.  As a result,
this study did not clearly identify a single trend (i.e., a fraction decreasing as a single power of the event number):
instead, it saw the superposition of several trends.
This approach was also severely limited by the small number of high-precision  mass distributions used ($\simeq 240$).

In this paper, we have outlined further tools to characterize a similar discrete sample of simulations: the effective
dimension and a local principal component analysis.   Both can be computed from a discrete sample.  For example, the
discrete cumulative log-likelihood distribution $P(<\delta \hat{L})$ found by comparing a synthetic data set to all
models can be approximated by a power law in the neighborhood of $\delta L \simeq 0$.  The exponent is the effective
dimension.  
Similarly, given a proposed (parameter-dependent) threshold on $\delta \hat{L}$, we can always find the set of models inside that 
contour, then perform a principal component analysis on the discrete collection of distributions $p$.  This
decomposition tells us what types of variations to expect in that (not always small) neighborhood.  
We have performed these calculations for test simulations and will provide detailed analysis in a subsequent publication.

\subsection{Measurement errors}
For simplicity, we have ignored the role of measurement error.   Gravitational wave detectors will be able to tightly
constrain some parameters, such as the ``chirp mass'' $(m_1 m_2)^{3/5}/(m_1+m_2)^{1/5}$, as these dramatically impact
the binary's rate of inspiral.  
In fact, the infinitesimal uncertainty in each chirp mass measurement will be far smaller than any expected features in the
binary chirp mass distribution.

Beyond this leading order dependence, however, very few parameters can be constrained precisely.   In direct opposition
to the chirp mass, the measurement accuracy for other parameters -- mass ratio,  spin magnitudes, and spin orientations
-- is often comparable to or larger than the expected width of these features: individual gravitational wave measurements
provide fairly little information \cite{1995PhRvD..52..848P}.
For the least-influential parameters, such as antisymmetric combinations of spins,  successive measurements will
only constrain our measurement uncertainty, not the underlying astrophysical distribution.
Finally, in several cases, systematic limitations in our ability to correctly model the long-lived signal from
inspiralling, precesing compact binaries prevent us from accurately reproducing source parameters \cite{gw-astro-PN-Comparison-AlessandraSathya2009}.    

For these parameters, more delicate comparisons of predicted distributions are warranted, that account for these
measurement errors.  
A detailed analysis of parameter estimation is far beyond the scope of this paper.
To leading order, however, we anticipate that parameter estimation uncertainties can be folded directly in to the
``predicted parameter distribution'' $p(x)$.  More concretely, if $p_{0}(x)$ is the posterior prediction of a particular
  binary evolution model and $K(x|x')$ is the conditional probability of recovering binary parameters $x$ given a
  measurement $x'$, then we assess the similarity of the (detector-dependent) predicted distributions $p(x)\equiv \int
  K(x|x')p_o(x)$.
As a first approximation, the conditional probability distribution can be estimated using standard Fisher matrix
techniques  \cite{1995PhRvD..52..848P}. 
The fraction of models consistent with observations can decrease so rapidly that it presents  severe computational
and data analysis challenges.  For example,  we need to distinguish between distributions
differing by $\Delta D \simeq 1/n^{d_{eff}/2-1}$.   To do this correctly, we must evaluate the likelihood correctly,
which in turn requires us to extract parameters for each event; to determine (data-analysis-strategy-dependent) selection biases for each type of binary;  et
cetera.   When the effective dimension is large, very small uncertainties about (or errors in) any stage in our data analysis pipeline can easily
contaminate the distribution comparisons described  above.

In fact, computational challenges also occur even in the absence of measurement error.   Model parameter distributions
in binary evolution are often sampled by Monte Carlo.   Because very high precision parameter distribution predictions are needed to
distinguish  neighboring models, the number of test binaries one must simulate  can scale as a prohibitive power of the
target likelihood accuracy.

\subsection{Reconstructing distributions from data?}
In this paper we have outlined a simple way to characterize model distribution differences, motivated by a simple 
maximum-likelihood statistic to differentate between proposed models.   This approach effectively characterizes
theoretical differences between predictions but would make suboptimal use of real gravitational wave data.   For
example, our method treats each detection at a point estimate with perfect confidence, not allowing for marginally
significant events.

Another approach to gravitational wave data analysis attempts to reconstruct the properties of the detected signal
distribution, without relying on any underlying models at all.   
This quasi-nonparametric approach has been applied both to hard astronomical data  astronomical problems
\cite{popsyn-constraint-StellarMassBHMassDistribution-Empirically-Farr2010} and the as-yet-hypothetical problem of
distinguishing binary mass distributions \cite{2010PhRvD..81h4029M}.    In this approach, one could employ the results of detailed Markov Chain Monte Carlo
studies (for example) to extract optimal posterior distributions from each measurement, then adjoin them to estimate the overall
population.  Such an approach could include marginal events and make full use of the available information per event.

These two techniques address fundamentally different questions, similar to nonparametric and modeled function
estimation.   In general if systematic biases are small and controlled, a modeled approach should permit more
precise constraints.   The method described in this paper naturally selects  model parameter distributions with greatest support (directly and via
principal components), as well as the plausible range of variations.    However, given the complexity of the model
space, nonparametric (unmodeled) distribution estimates provide a vital corroborating test of model-fit results.  We
will address the problem of comparing modeled and unmodeled parameter estimates in a subsequent paper.

\section{Conclusions}
The astrophysical interpretation of gravitational wave data faces a common dilemma: observations produce a small data
set (there, compact object numbers and parameters) which need to be compared to  expensive, complex predictions.  In
this paper, we introduce methods to facilitate the identification and interpretation of these kinds of comparisons,
using gravitational wave chirp mass measurements as an example.   
To better understand the model space, we suggest principal component analysis on small model subsets, to
nonparametrically identify local
(highly variable) degrees of freedom that impact the mass distribution.  
We also propose a new family of (local) coordinates on the model space.  Rather than choose simulation input parameters, these coordinates are naturally adapted to the
physically identifiable degrees of freedom, as characterized by the range of predicted parameter distributions.  
Using these coordinates, we can transparently address how much information each successive observation provides.

Previous studies suggest that predicted parameter distributions have a range of differences, from dramatic to minute.
For suitable coordinates, one population parameter $\lambda_1$ may have a dramatic effect on the population (e.g., the
location of the dominant peak) while others may have far less notable impact (e.g., the location of a very small
feature).   Evidently, only a handful of observations are needed to identify and measure the first variation; many
measurements are needed to recognize the second variation exists.  
Our techniques correctly identify that these scales exist; determine how many events are needed to resolve them; and,
for each number of events, characterize the relevant number of degrees of freedom.   
In particular, we can predict how well each successive event distinguishes a model from its neighbors: the fraction of
consistent models decreases as a power $n^{-d_{eff}/2}$ for $d_{eff}$ an effective dimension characterizing the local
model space.
This fraction can decrease very rapidly if, as expected, the predicted mass and spin distributions differ significantly
between realizations.   
Past a certain point, data analysis pipelines and progenitor model simulations must take great care to insure they
provide sufficiently high-accuracy parameter distributions, to best exploit the  available information.
At the other extreme, our principal component technique lets us identify  when the model space
is large and a purely phenomenological approach is needed.
Our methods can incorporate estimates for experimental uncertainty.  
In short, we present a concrete way to deal simultaneously with experimental uncertainty and model complexity: we let
the model space itself identify what is important.

 Our discussion is  broadly applicable to experiments
where on the one hand experiments measure several properties of each event but on the other hand theory cannot produce comparable multivariate
distributions without significant computational cost.  Our method is statistically well-posed;  applies equally well to
coarse or fine sampling of the model space; can employ as many or few dimensions as needed; can be easily modified to
account for experimental error; and asymptotes to a maximally informative Fisher matrix analysis in the
limit of small statistical and systematic errors.

In a subsequent publication, we will use concrete binary evolution codes like \texttt{BSE} and \texttt{StarTrack} to
characterize the relevant degrees of freedom for binary compact objects.  
Based on their concrete example, we will assess the computational requirements for future simulations needed to 
 take full advantage of future gravitational wave observations by advanced ground-based detectors like advanced LIGO,
 advanced Virgo, and the Einstein Telescope.
In another publication we will use perturbative calculations to assess the limits of this approach, showing how third-generation
detectors can provide extremely high-precision estimates for binary evolution parameters.

\appendix
\section{Principal component analysis}
The natural eigendirections associated with a data set can be characterized by principal component analysis
\cite{mm-statistics-nonparametric-timeseries-RamsaySilverman}.  In one sense, principal component analysis corresponds
to finding the best set of $N$ orthonormal basis functions $\phi_k(x)$ with which to approximate a set of $N_{samp}$ functions $\delta
p_A$, in that the sample-summed error is smallest:
\begin{eqnarray}
\text{global error} = \sum_A || \delta p_A - \delta \hat{p}_A||^2
\end{eqnarray}
In this expression $\delta \hat{p}_A$ is the projection of $\delta p_A$ onto the space spanned by the $N$ basis
functions, and therefore the optimal estimate for it in that space.  
Each orthonormal basis function contributes independently to the global squared error.  
This variational problem immediately leads to an eigenvalue problem for the correlation operator ${\cal C}$ [Eq. (\ref{eq:def:Corr})].

\hidetosubmit{
\section{Fluctuations in log likelihood(*)}
\editremark{may put this in?}
}

\acknowledgements
ROS is currently supported by NSF award PHY-0970074, the Bradley Program Fellowship, and the UWM Research
Growth Initiative.  ROS was also supported by NSF award PHY 06 -53462 and the Center for
  Gravitational Wave Physics, where this work  commenced.

\bibliography{%
LIGO-publications,%
mm-statistics,%
gr-tests-extensions-models,%
observations-bh-stellar,%
gw-astronomy,gw-astronomy-mergers,gw-astronomy-mergers-approximations,gw-astronomy-mergers-nr,gw-astronomy-detection-statistics,%
gw-astronomy-stochastic-mergers,%
gw-astronomy-detection,%
popsyn,popsyn_gw-merger-rates,popsyn-constraints,%
short-grb,short-grb-mergermodel,%
astrophysics-stellar-dynamics-theory,astrophysics-stellar-dynamics-theory-withStellarAndBinaryEvolution,%
structureformation-smbh-binaries}

\end{document}